\documentstyle[preprint,floats,aps,psfig]{revtex}
\tightenlines
\begin{document}

\draft

\title{Nucleon-nucleon elastic scattering to 3~GeV}
\author{Richard A. Arndt\thanks{arndtra@enterprise.phys.vt.edu}, 
        Igor I. Strakovsky\thanks{igor@gwu.edu}, 
        and Ron L. Workman\thanks{rworkman@gwu.edu}}
\address{Center for Nuclear Studies and Department of Physics \\
        The George Washington University Washington, DC 20052}

\draft
\date{\today}
\maketitle

\begin{abstract}
A partial-wave analysis (PWA) of $NN$ elastic 
scattering data has been completed.  This analysis 
covers an expanded energy range, from threshold to 
a laboratory kinetic energy of 3~GeV, in order to 
include recent elastic $pp$ polarized scattering 
measurements performed at SATURNE~II. Results of 
the energy-dependent fit are compared with 
single-energy solutions and Saclay amplitudes
obtained via the direct-reconstruction approach.  We also 
comment on the status of $\epsilon_1$ in the 
low-energy region.\\
\end{abstract}
\vspace*{0.5in}

\pacs{PACS numbers: 11.80.Et, 13.75.Cs, 25.40.Cm, 25.40.Dn}

\narrowtext
\section{Introduction}
\label{sec:intro}

This analysis of elastic nucleon-nucleon scattering 
data updates our previous analyses to 1.6~GeV 
\cite{vpi94} and 2.5~GeV \cite{vpi97} in the 
laboratory kinetic energy.  The present analysis 
has been extended to 3~GeV in order to include both 
the elastic $pp$ polarized measurements \cite{al98} 
$-$ \cite{ba992} from SATURNE~II at Saclay, and $pp$ 
differential cross sections measured \cite{edda} by 
the EDDA collaboration using the cooler synchrotron 
at COSY.  A detailed description of our database is 
given in Section~II.

As discussed in Ref.\cite{vpi97}, the region beyond 
2~GeV is interesting for several reasons.  These 
include the suggestion of a narrow dibaryon resonance, 
corresponding to a center-of-mass energy of 2.7~GeV. 
Near this energy, a sharp structure was found in
preliminary $A_{yy}$ measurements\cite{ball} and 
this was taken as support for such a resonance.
The authors of Ref.~\cite{edda} considered this
possibility but found no evidence in their 
differential cross section measurements. No 
significant anomaly was seen in the angular and
energy dependence of detailed 
analyzing power\cite{ar97} and
correlated spin measurement at Saclay\cite{al995p}.

The possible onset of behavior suggested by 
dimensional counting at high energy and fixed 
center-of-mass angle is also intriguing\cite{brodsky,farrar}.
In Ref.\cite{vpi97}, we noted that $d\sigma / dt$
appeared to be approaching $s^{-10}$, as expected within
perturbative QCD. Thus, an extended energy range is
needed to verify this trend. An extension of the $np$ 
analysis beyond 1.3~GeV would also benefit those 
studying the two-body photodisintegration of the 
deuteron at Jefferson Lab, 
which shows an interesting scaling behavior 
at some scattering angles \cite{mark,jlab}.  
Unfortunately, the $np$ data base remains too sparse 
to support a reliable analysis beyond 1.3~GeV.

In the present work, we have focused mainly on the 
influence of new polarization data at higher energies, 
and on the behavior of $\epsilon_1$ at low energies.  
The Saclay group has recently performed a 
single-energy phase-shift 
analysis of elastic $pp$ scattering data to 2.7 GeV
and $np$ elastic scattering 
data to 1.1 GeV\cite{sac2}.  
In this study, a second set of 
amplitudes\cite{sac2,sac1} was obtained 
through a direct reconstruction of the scattering 
amplitudes at fixed energies and angles.  We have 
compared our results to these and, in some cases, find
evidence (complimentary to that given by the Saclay 
group) for non-uniqueness at higher energies.  At lower 
energies, where the behavior of $\epsilon_1$ has been 
a source of controversy, we compare our results to 
those of several other groups and suggest there is 
little evidence for an anomalously large tensor 
interaction. Results of our analyses are displayed in
Section III. In Section IV, we summarize our findings
and conclusions.

\section{Database}
\label{sec:data}

Our two previous $pp$ scattering analyses 
\cite{vpi94,vpi97} extended to 1.6 and 2.5~GeV,
respectively.  In both cases, the associated $np$ 
analysis was restricted to 1.3~GeV.  The present data 
base\cite{said} is considerably larger due both to an 
expanded energy range for the $pp$ system and the 
addition of new data at lower energies.  The full 
data base has increased by 30\% since the publication 
of Ref.\cite{vpi97}, and is about 70\% larger than the 
set available for the analysis of Ref.\cite{vpi94}. 
(The total data base has doubled over the last decade 
(see Table~I).)  The distribution of recent (post-1997) 
$pp$ and $np$ data is given in Fig.~1.  

In the full data base, one will occasionally find 
experiments which give conflicting results.  Some of
these have been excluded from our fits.  We have, 
however, retained all available data sets so that 
comparisons can be made through our on-line facility
\cite{said}.  Below, we list recent additions to our 
data base.  Some of the data listed as new were available, 
in unpublished form, at the time 
of our previous analysis\cite{vpi97}. 
A complete description  of the data base and those data 
not included in our analyses is available from the 
authors. 

Two thirds of the 4802 new $pp$ polarization data 
were produced at Saclay using the SATURNE~II 
accelerator \cite{al98} $-$ \cite{ba992}.  These 
measurements of 9 spin-dependent quantities 
have increased our data base by a 
factor of two over the energy range from 1.6 to  
2.5~GeV and accounted for a third of the data from
2.5 to 3.0~GeV.  Many of the new $pp$ polarization 
measurements below 450~MeV were made at the Indiana 
cooler (A$_y$, A$_{yy}$, A$_{xx}$, A$_{zz}$, and 
A$_{xz}$) \cite{ha97} $-$ \cite{wi99}.  Also, in this 
energy range are 11 new unpolarized cross sections 
(at 398~MeV) measured at the Osaka facility 
\cite{ya99}.

The $np$ data base has not increased significantly.  
As a result, we have not extended the range of our 
analyses for the $I = 0$ system beyond 1.3 GeV.  
The Geneva group\cite{ah98}, working at PSI,
has provided 247 new $np$ spin 
measurements. From this source, we have also 
added spin observables A$_y$, A$_t$, D$_t$, and 
R$_t$ from 260 to 538~MeV \cite{ah98}.  About 60\% 
of the recent SATURN~II $np$ polarized measurements
fall within our energy range; the full range extends 
from about 1.1 to 2.4~GeV (182 data points) \cite{le99}.  
A few measurements of $\Delta\sigma _L$ and 
$\Delta\sigma_T$ were produced by TUNL \cite{ra99} and Charles 
University at Prague \cite{br97}.  A single 
measurement of D$_t$, at 16~MeV, was provided 
\cite{cl98} by the ISKP cyclotron at Bonn. 
We have added 12 unpolarized measurements, 
between 29 and 73~MeV, from 
the Louvain-la-Neuve Cyclotron \cite{be97}.
Finally, we have retained in the analysis
a set of 82 total cross section measurements, 
between 4 and 231~MeV \cite{li80}, which had
earlier been removed in order to have a low-energy
data base identical to that used by the Nijmegen
group\cite{nijpwa}.

\section{Partial-Wave Analyses}
\label{sec:pwa}

Fits to the expanded data base and extended energy 
range were found to be possible within the formalism 
used and described in our previous analyses
\cite{vpi94,vpi97}.  Both energy-dependent and 
single-energy solutions were obtained from fits to 
the combined $pp$ and $np$ data bases to 1.3 GeV, and
from fits to the $pp$ data base alone from 1.3 to
3 GeV.  In Table~II, 
we compare the energy-dependent and single-energy 
results over the energy bins used in the 
single-energy analyses.  Also listed are the number 
of parameters varied in each single-energy solution.  
A total of 147 parameters were varied in the 
energy-dependent analysis (SP00).

Our single-energy and energy-dependent results for 
the isovector and isoscalar partial-wave amplitudes 
are displayed in Figs.~2 and 3.  Here, we also 
compare with our previous fit (SM97)\cite{vpi97} 
and a much older fit (FA91)\cite{vpi92}. Partial
waves with $J < 6$ are displayed, whereas the
analysis fitted waves up to $J=7$.
In most cases, SP00 and SM97 are in good 
agreement.  Somewhat larger changes are seen 
in comparisons with FA91.  Differences are 
generally largest, as one would expect, near the 
energy upper limits for the various solutions and 
in the smaller partial waves. Figs.~2 and 3 therefore
show that a doubling of the data base has led to a 
refinement of the amplitudes, but has not
required a dramatic change in their behavior.

Single-energy solutions were produced up to 
2825~MeV (for $pp$ scattering).  In these fits, 
initial values for the partial-wave amplitudes and 
their (fixed) energy derivatives were obtained 
from the energy-dependent solution.  A comparison 
of global and single-energy solutions then serves 
as a check for structures that could have been 
``smoothed over'' in the energy-dependent analysis.
However, structures with widths less than 10 MeV
would be very difficult to detect.

In order to ascertain that the extension to 3~GeV 
(1.3~GeV for $np$) did not seriously degrade our 
low-energy results, a 0 $-$ 400~MeV fit was also 
developed.  The resultant solution, SP40, used 30 
$I = 1$ and 27 $I = 0$ variable parameters to give 
a $\chi^2$/datum of 4398/3454~($pp$), and 
5415/3831~($np$).  The global fit, SP00, produced, 
for the same energy range, a $\chi^2$/datum of 
4593/3454~($pp$) and 5371/3831~($np$).  We 
consider this quite reasonable given that the number 
of variable parameters per datum is nearly three 
times larger for SP40 than for SP00.  

In Figs.~4 and 5, we compare our results with the 
Saclay single-energy analyses \cite{sac2} for 
isovector waves below 2.7 GeV and 
isoscalar waves below 1.1 GeV. In the isoscalar 
case, the agreement is quite good, given the 
overall scatter of single-energy fits around the 
global result.  More substantial differences are 
seen in the Saclay results for $I = 1$ partial 
waves above 1 GeV. 

A possible explanation for this discrepancy is 
given in the recent Saclay amplitude-reconstruction 
analysis\cite{sac2,sac1}.  
In Fig.~6, we compare the Saclay results 
to curves generated from our single-energy and 
energy-dependent solutions.  (A similar comparison 
was made in the $I = 0$ case, with good overall 
agreement between the three solutions.)  Here, we 
are using the notation of Ref.\cite{sac2} and write 
the scattering matrix, $M$, as
\begin{eqnarray}
M(\vec k_f , \vec k_i ) & = & {1\over 2} [ (a + b) +
       (a - b)~(\vec \sigma_1 \cdot \vec n)
              ~(\vec \sigma_2 \cdot \vec n)
   +   (c + d)~(\vec \sigma_1 \cdot \vec m )
              ~(\vec \sigma_2 \cdot \vec m ) \nonumber \\
 & + & (c - d)~(\vec \sigma_1 \cdot \vec l )
              ~(\vec \sigma_2 \cdot \vec l )
   +         e~(\vec \sigma_1 +\vec \sigma_2 )\cdot \vec n ] ,
\end{eqnarray}
where $\vec k_f$ and $ \vec k_i$ are the scattered 
and incident momenta in the center-of-mass system, 
and
\[ \vec n = { { \vec k_i \times \vec k_f }
        \over {|\vec k_i \times \vec k_f |} }
  \; , \;
   \vec l = { { \vec k_i + \vec k_f  }
        \over {|\vec k_i + \vec k_f | } }
  \; , \;
   \vec m = { { \vec k_f - \vec k_i  }
        \over {|\vec k_f - \vec k_i | } } .
 \]
In Refs. \cite{sac2,sac1}, multiple solutions were found 
at most energy-angle points.  The $pp$ amplitudes are plotted 
together in Fig.~6 where we can see that, in some 
cases, our single-energy results favor one branch,
while the energy-dependent fit follows another.
This feature was also evident in Ref. \cite{sac2}, 
where it was shown that the Saclay partial-wave 
analyses followed a branch different from our 
preliminary energy-dependent fits. 

Finally, in Fig.~7, we return to the low-energy 
region which has been controversial mainly due to 
recent determinations of $\epsilon_1$.  In Ref. 
\cite{ra99}, the trend of recent determinations 
was used to argue for an $NN$ tensor interaction 
stronger than predicted by all meson-exchange-based 
potential models, and in conflict with values found 
in both the Nijmegen and VPI partial-wave analyses. 
This trend is absent in our figure, where we have 
compared two of our energy-dependent fits, 
SP00 and SP40, and the 
Nijmegen potential, to a selection of recent 
single-energy fits.  Clearly there is considerable 
scatter in the single-energy results.  However, 
given this variation, the overall agreement with 
energy-dependent fits is quite good. 

\section{Conclusions and Future Prospects}
\label{sec:conc}

In our previous analysis \cite{vpi97}, an extension of 
the energy range for $pp$ elastic scattering, from 1.6 
to 2.5~GeV, was mainly motivated by the addition of 
precise new (unpolarized) cross section measurements 
from the EDDA collaboration \cite{edda}.  Given the 
sparse polarization data in this region, the fit was 
expected to change significantly with the addition of 
Saclay\cite{al98} $-$ \cite{ba992} 
and future COSY (polarized) measurements.  It 
is therefore somewhat surprising how little our new 
solution (SP00) has changed from SM97 \cite{vpi97}.

We have seen that the $I = 0$ amplitudes are generally 
in good agreement with the Saclay results.  This holds 
true in the low-energy region as well, the Saclay value 
for $\epsilon_1$ being consistent with our result.  
However, the agreement for $pp$ ($I = 1$) amplitudes 
above 1~GeV is less impressive. 

As mentioned above, this difference in partial-wave 
solutions may be a reflection of the non-uniqueness 
seen in the Saclay amplitude reconstruction.  The 
selection of data included in these analyses could 
also be a factor. Clearly, this serves as further 
motivation for the polarization measurements being
performed at COSY \cite{cosy} and JINR \cite{jinr}. 

\acknowledgments

The authors express their gratitude to J. Ball, Z. 
Dolezal, F. Dohrmann, W. Haeberli, H. O. Klages, C. 
Lechanoine-Leluc, F. Lehar, B. Lorentz, H.-O. Meyer, 
N. Olsson, B. von Przewoski, H. M. Spinka, W. Tornow, 
S. W. Wissink, and K. Yasuda for providing experimental 
data prior to publication or for clarification of 
information already published.  Also, we thank F. Lehar 
for fruitful discussions and PWA predictions.  This 
work was supported in part by the U.~S. Department of 
Energy Grant DE--FG02--99ER41110.  The authors gratefully 
acknowledges a contract from Jefferson Lab under which 
this work was done.  Jefferson Lab is operated by the 
Southeastern Universities Research Association under the 
U.~S.~Department of Energy Contract DE--AC05--84ER40150.

\eject


\eject

{\Large\bf FIGURE CAPTIONS}\\
\newcounter{fig}
\begin{list} {Figure \arabic{fig}.}
{\usecounter{fig}\setlength{\rightmargin}{\leftmargin}}
\item
{Energy-angle distribution of recent 
(post-1997) (a) $pp$ and (b) $np$ data.  
The $pp$ data contribution below 3000~MeV 
is 30\% and most new data are 
A$_y$~(47\%) or A$_{yy}$~(29\%). The $np$ 
data contribution below 1300~MeV is 6\% 
and most new data are 
A$_y$~(24\%).  Total cross sections are 
plotted at zero degrees.}
\item
{Isovector partial-wave amplitudes from 
0 to 3~GeV in the proton kinetic energy.  
Solid (dashed) curves give the real 
(imaginary) parts of amplitudes 
corresponding to the SP00 solution.  
The real (imaginary) parts of 
single-energy solutions are plotted as 
filled (open) circles.  The 
SM97 solution \protect\cite{vpi97} is 
plotted with long dash-dotted (real 
part) and short dash-dotted (imaginary 
part) lines.  FA91 solution 
\protect\cite{vpi92} is shown by dashed
lines.  The dotted curve gives the unitarity 
limit \hbox{Im$T$ - $T^2$ - $T_{sf}^2$} from 
SP00, where $T_{sf}$ is the spin-flip 
amplitude.  All amplitudes are 
dimensionless.}
\item
{Isoscalar partial-wave amplitudes from 
0 to 1.2~GeV.  Notation as in Fig.~2.}
\item
{Phase-shift parameters for isovector
partial-wave amplitudes from 0 to 
3000~MeV.  The SP00 and SM97 
\protect\cite{vpi97} solutions are 
plotted as solid and dash-dotted 
curves, respectively.  Our single-energy
solutions and those from 
Saclay \protect\cite{sac2,sac1}) 
are given by filled and open circles,
respectively.}
\item
{Phase-shift parameters for isoscalar
partial-wave amplitudes from 0 to 
1200~MeV.  Notation as in Fig.~6.}
\item
{Direct-reconstruction scattering 
amplitudes at 
(a)~1.80~GeV, (b)~2.10~GeV, 
(c)~2.40~GeV, and (d)~2.70~GeV.  The real 
(imaginary) parts of amplitudes $a$ to 
$e$ \protect\cite{sac2} are shown 
in $\sqrt{mb/sr}$ as a function of the 
c.~m. scattering angle and plotted as 
filled (open)  circles.  Our SP00 
(single-energy) solution is 
plotted with solid (dashed) lines.}
\item
{Summary of analyses giving 
$\epsilon _1$ in the energy range up 
to 80~MeV.  The solid (dashed) curve gives
our SP00 (SP40) PWA results.  Nijmegen 
potential results \protect\cite{nijm} 
are plotted as a dash-dotted line.  Filled 
circles (diamonds) give GW 
(Saclay \protect\cite{sac2}) single-energy 
PWA results.  Open 
squares denote the single-energy PWA from PSI 
\protect\cite{he93}.  Other results are 
from TUNL (star) \protect\cite{ra99},
Bonn (filled box) \protect\cite{oc91,cl98}, 
Prague (open diamond) \protect\cite{br97}, 
Erlangen (open circle) \protect\cite{sc88}, 
PSI (filled triangle) \protect\cite{ha92}, 
and Karlsruhe (open triangle) \protect\cite{do89}.
}
\end{list}
\eject

Table~I. Comparison of present (SP00 and SP40) 
and previous (SM97, SM94, VZ40, FA91, SM86, and 
SP82) energy-dependent partial-wave analyses.  
The $\chi ^2$ values for the previous solutions 
correspond to our published results 
(\cite{vpi94}, \cite{vpi97}, and \cite{vpi92} 
$-$ \cite{vpi83}).
\vskip 10pt
\centerline{
\vbox{\offinterlineskip
\hrule
\hrule
\halign{\hfill#\hfill&\qquad\hfill#\hfill&\qquad\hfill#\hfill
&\qquad\hfill#\hfill&\qquad\hfill#\hfill&\qquad\hfill#\hfill\cr
\noalign{\vskip 6pt} %
Solution&Range~(MeV)&$\chi^2$/$pp$~data&Range~(MeV)&$\chi^2$/$np$~data&
Ref. \cr
\noalign{\vskip 6pt}
\noalign{\hrule}
\noalign{\vskip 10pt}
SP00 & $0 -3000$ &  36617/21796 & $0 -1300$ & 18693/11472 & Present  \cr
\noalign{\vskip 6pt}
SP00 &$(0 -2500)$&  34277/20947 & $0 -1300$ & 17693/11330 & Present  \cr
\noalign{\vskip 6pt}
SP00 &$(0 -1600)$&  23927/15766 & $0 -1300$ & 17693/11330 & Present  \cr
\noalign{\vskip 6pt}
SP00 &$(0  -400)$&   4593/ 3454 &$(0  -400)$&  5371/ 3831 & Present  \cr
\noalign{\vskip 6pt}
SP40 & $0 - 400$ &   4398/ 3454 & $0 - 400$ &  5415/ 3831 & Present  \cr
\noalign{\vskip 6pt}
SM97 & $0 -2500$ &  28686/16994 & $0 -1300$ & 17437/10854 & \cite{vpi97} \cr
\noalign{\vskip 6pt}
SM94 & $0 -1600$ &  22371/12838 & $0 -1300$ & 17516/10918 & \cite{vpi94} \cr
\noalign{\vskip 6pt}
VZ40 & $0  -400$ &   3098/2170  & $0  -400$ &  4595/3367  & \cite{vpi94} \cr
\noalign{\vskip 6pt}
FA91 & $0 -1600$ &  20600/11880 & $0 -1100$ & 13711/7572  & \cite{vpi92} \cr
\noalign{\vskip 6pt}
SM86 & $0 -1200$ &  11900/7223  & $0 -1100$ &  8871/5474  & \cite{vpi87} \cr
\noalign{\vskip 6pt}
SP82 & $0 -1200$ &   9199/5207  & $0 -1100$ &  9103/5283  & \cite{vpi83} \cr
\noalign{\vskip 10pt}}
\hrule}}
\vfill
\eject
Table~II. Comparison of the single-energy (SES) 
and energy-dependent (SP00) fits to $pp$ and 
$np$ data.  Values of $\chi^2$ are given for 
the single-energy and SP00 fits 
(evaluated over the same 
energy bins).  Also listed is the number of 
parameters varied in each single-energy 
solution.

\vskip 10pt
\centerline{
\vbox{\offinterlineskip
\hrule
\hrule
\halign{\hfill#\hfill&\qquad\hfill#\hfill&\qquad\hfill#\hfill
 &\qquad\hfill#\hfill \cr
\noalign{\vskip 6pt}
Energy Range (MeV) & $\chi^2$ SES(SP00)/$pp$~data
 & $\chi^2$ SES(SP00)/$np$~data & Parameters  \cr
\noalign{\vskip 6pt}
\noalign{\hrule}
\noalign{\vskip 10pt}
   4-6    &     22(39)/28    &     78(83)/63   &   6  \cr
\noalign{\vskip 6pt}
   7-12   &    84(134)/88    &   254(333)/101  &   6  \cr
\noalign{\vskip 6pt}
  11-19   &     17(47)/27    &   205(455)/247  &   8  \cr
\noalign{\vskip 6pt}
  19-30   &   123(268)/114   &   292(321)/316  &   8  \cr
\noalign{\vskip 6pt}
  32-67   &   282(354)/224   &   809(879)/548  &  10  \cr
\noalign{\vskip 6pt}
  60-90   &     48(63)/72    &   514(629)/355  &  10  \cr
\noalign{\vskip 6pt}
  80-120  &   152(156)/154   &   465(453)/382  &  10  \cr
\noalign{\vskip 6pt}
 125-174  &   313(336)/287   &   603(653)/333  &  11  \cr
\noalign{\vskip 6pt}
 175-225  &   494(542)/435   &   701(734)/504  &  13  \cr
\noalign{\vskip 6pt}
 225-270  &   222(246)/228   &   299(345)/278  &  13  \cr
\noalign{\vskip 6pt}
 276-325  &   771(802)/740   &   628(680)/564  &  17  \cr
\noalign{\vskip 6pt}
 325-375  &   460(474)/406   &   416(460)/353  &  17  \cr
\noalign{\vskip 6pt}
 375-425  &   738(758)/607   &   804(870)/599  &  17  \cr
\noalign{\vskip 6pt}
 425-475  & 1055(1156)/803   &   828(870)/682  &  18  \cr
\noalign{\vskip 6pt}
 475-525  & 1311(1565)/1081  & 1248(1404)/787  &  30  \cr
\noalign{\vskip 6pt}
 525-575  &   858(956)/754   &   672(694)/488  &  31  \cr
\noalign{\vskip 6pt}
 575-625  & 1039(1112)/760   &   423(484)/367  &  34  \cr
\noalign{\vskip 6pt}
 625-675  &   908(842)/773   & 1270(1611)/873  &  36  \cr
\noalign{\vskip 6pt}
 675-725  &   860(923)/797   &   404(468)/386  &  37  \cr
\noalign{\vskip 6pt}
 725-775  & 1007(1311)/827   &   518(556)/381  &  37  \cr
\noalign{\vskip 6pt}
 775-824  & 1690(1840)/1301  & 1550(1861)/948  &  38  \cr
\noalign{\vskip 6pt}
 827-874  & 1155(1330)/914   &   388(467)/365  &  39  \cr
\noalign{\vskip 6pt}
 876-924  &  342(475)/389    &   752(905)/625  &  41  \cr
\noalign{\vskip 6pt}
 926-974  &   762(992)/679   &   363(512)/353  &  43  \cr
\noalign{\vskip 10pt}}
\hrule
\hrule}}
\vfill

\eject

Table~II. (continued)

\vskip 10pt
\centerline{
\vbox{\offinterlineskip
\hrule
\hrule
\halign{\hfill#\hfill&\qquad\hfill#\hfill&\qquad\hfill#\hfill
 &\qquad\hfill#\hfill \cr
\noalign{\vskip 6pt}
Energy Range (MeV) & $\chi^2$ SES(SP00)/$pp$ data
 & $\chi^2$ SES(SM97)/$np$ data & Parameters  \cr
\noalign{\vskip 6pt}
\noalign{\hrule}
\noalign{\vskip 10pt}
 976-1020 &  917(1177)/708   &   284(425)/328  &  43  \cr
\noalign{\vskip 6pt}
1078-1125 &  815(1128)/573   &   519(846)/427  &  47  \cr
\noalign{\vskip 6pt}
1261-1299 &  691(1006)/505   &       $---$     &  30  \cr
\noalign{\vskip 6pt}
1481-1521 &   140(307)/149   &       $---$     &  30  \cr
\noalign{\vskip 6pt}
1590-1656 &   505(892)/460   &       $---$     &  31  \cr
\noalign{\vskip 6pt}
1685-1724 &   174(309)/116   &       $---$     &  31  \cr
\noalign{\vskip 6pt}
1778-1818 &  625(1097)/506   &       $---$     &  33  \cr
\noalign{\vskip 6pt}
1929-1975 &   377(463)/366   &       $---$     &  33  \cr
\noalign{\vskip 6pt}
2065-2120 & 1173(1938)/829   &       $---$     &  33  \cr
\noalign{\vskip 6pt}
2175-2225 & 1476(2046)/758   &       $---$     &  33  \cr
\noalign{\vskip 6pt}
2330-2470 & 1013(1808)/713   &       $---$     &  33  \cr
\noalign{\vskip 6pt}
2500-2600 &   250(523)/311   &       $---$     &  33  \cr
\noalign{\vskip 6pt}
2642-2714 &  302(1016)/307   &       $---$     &  33  \cr
\noalign{\vskip 6pt}
2792-2869 &   148(405)/153   &       $---$     &  33  \cr
\noalign{\vskip 10pt}}
\hrule
\hrule}}
\vfill
\eject

\end{document}